\documentclass[11pt,twoside]{article}
\usepackage{./asp2010}
\usepackage{times}
\usepackage{natbib} 
\usepackage{amssymb}
\usepackage{textcomp}
\usepackage[normalem]{ulem}
\resetcounters

\begin{document}

\title{Clues to the Formation of Lenticular Galaxies Using Spectroscopic Bulge--Disk Decomposition}
\author{Evelyn~J.~Johnston,$^1$ Alfonso~Arag\'on-Salamanca,$^1$ Michael~R.~Merrif\mbox{}ield$^1$ and Alejandro G. Bedregal$^{2}$
\affil{$^1$School of Physics and Astronomy, University of Nottingham, University Park, Nottingham, NG7 2RD, UK}
\affil{$^2$Department of Physics and Astronomy, Tufts University, Medford, MA 02155, USA}}

\begin{abstract}
Lenticular galaxies have long been thought of as evolved spirals, but the 
processes involved to quench the star formation are still unclear. By 
studying the individual star formation histories of the bulges and disks of 
lenticulars, it is possible to look for clues to the processes that 
triggered their transformation from spirals. To accomplish this feat, 
we present a new method for spectroscopic 
bulge--disk decomposition, in which a long-slit spectrum is decomposed 
into two one-dimensional spectra representing purely the bulge and 
disk light. We present preliminary results from applying this method 
to lenticular galaxies in the Virgo and Fornax Clusters, in which we 
show that the most recent star formation activity in these galaxies 
occurred within the bulges. We also find that the bulges are in general more 
Fe-enriched than the disks of the same galaxy, and that this enrichment 
grows stronger as the age of the bulge becomes younger. These results point 
towards a scenario where the star formation in the disks of spiral galaxies 
are quenched, followed by a burst of star formation in the central regions
from the gas that has been funnelled inwards through the disk.

\end{abstract}

\section{Introduction}
Spirals and lenticulars (S0s) lie next to each other on the Hubble Sequence, 
where both display disky morphologies with young and old stellar populations 
respectively. As a result, S0s are often seen as a possible endpoint of 
the evolution of spiral galaxies. This idea is backed up by studies 
such as \citet{Dressler_1980}, where the fraction of spirals was found 
to decrease in higher density environments while that of S0s increased, 
suggesting a direct link between the two morphologies and their environment. 
Later studies by \citet{Dressler_1997}, \citet{Fasano_2000} and \citet{Desai_2007} 
have also shown a relationship with redshift, where the fraction of 
S0s increases at lower redshifts while that of spirals decreases with the 
transition occurring between 
2 and 5 billion years ago. These results make S0s in rich clusters the ideal 
targets with which to study the transformation from spirals.

Many processes have been proposed to explain the transformation of spirals 
to S0s, most of which focus on the truncation of star formation in the disk 
followed by passive evolution as the galaxy fades into an S0. Examples of 
such processes include ram pressure stripping, starvation, tidal stripping by 
galaxy harassment, and starbursts 
triggered by unequal mass galaxy mergers and galaxy--cluster
interactions, each of which 
would affect the bulge and disk in different ways. Therefore, it should be 
possible to determine the transformation process by studying the 
star formation histories of the bulge and disk independently.

\section{Spectroscopic Bulge--Disk Decomposition}
In order to study the star formation histories of the bulge and disk, the 
light was first separated into individual bulge and disk 
spectra by spectroscopic bulge--disk decomposition \citep{Johnston_2012}.
In brief, this procedure involves taking the light profile of the galaxy at each 
wavelength in the spectrum, and fitting a bulge and disk light 
profile to this spatial distribution in the same way as for one-dimensional 
photometric bulge--disk decomposition. In each case, a simple S\'ersic bulge 
plus exponential disk profile was used for the decomposition. Having obtained 
the bulge and disk parameters at each wavelength from the best fit to the light 
profile, the total light from each component at that wavelength was calculated 
through integration and tabulated against wavelength to produce high-quality 
one-dimensional bulge and disk spectra.

This method was applied to a sample of 30 S0s from the Virgo and Fornax Clusters 
with inclinations above 40\textdegree. The spectra were observed in long-slit 
mode on Gemini-GMOS (Virgo) and VLT-FORS2 (Fornax). The observations covered a 
magnitude range of \mbox{ $-22.3 < M_{B} < -17.3$}, and a wavelength range of 
\mbox{$4100 < \lambda < 5900\ \AA$}. Exposure times were typically between 
2 and 3.5 hours to ensure a S/N above 50 at the peak of the spectrum. Of these 
30 galaxies, 18 could be decomposed successfully with the simple bulge plus 
disk model used in this study. The remaining galaxies could not be decomposed due 
to more complicated light profiles that could not be modelled in this way, or 
the presence of very compact bulges that could not be fitted reliably.

\section{Star Formation Histories of the Bulge and Disk}
The decomposed bulge and disk spectra hold clues to their star formation 
histories within their absorption line strengths. 
Hydrogen absorption lines are often used as an indicator of the age of 
the stellar population dominating the light from a galaxy, which in turn 
tells us how long ago the last star formation event happened. Similarly, 
magnesium and iron lines can be used to measure the metallicity of the 
stellar population, and thus provide clues as to the earlier star 
formation history of the galaxy. 

\begin{figure}[t]
\plottwo{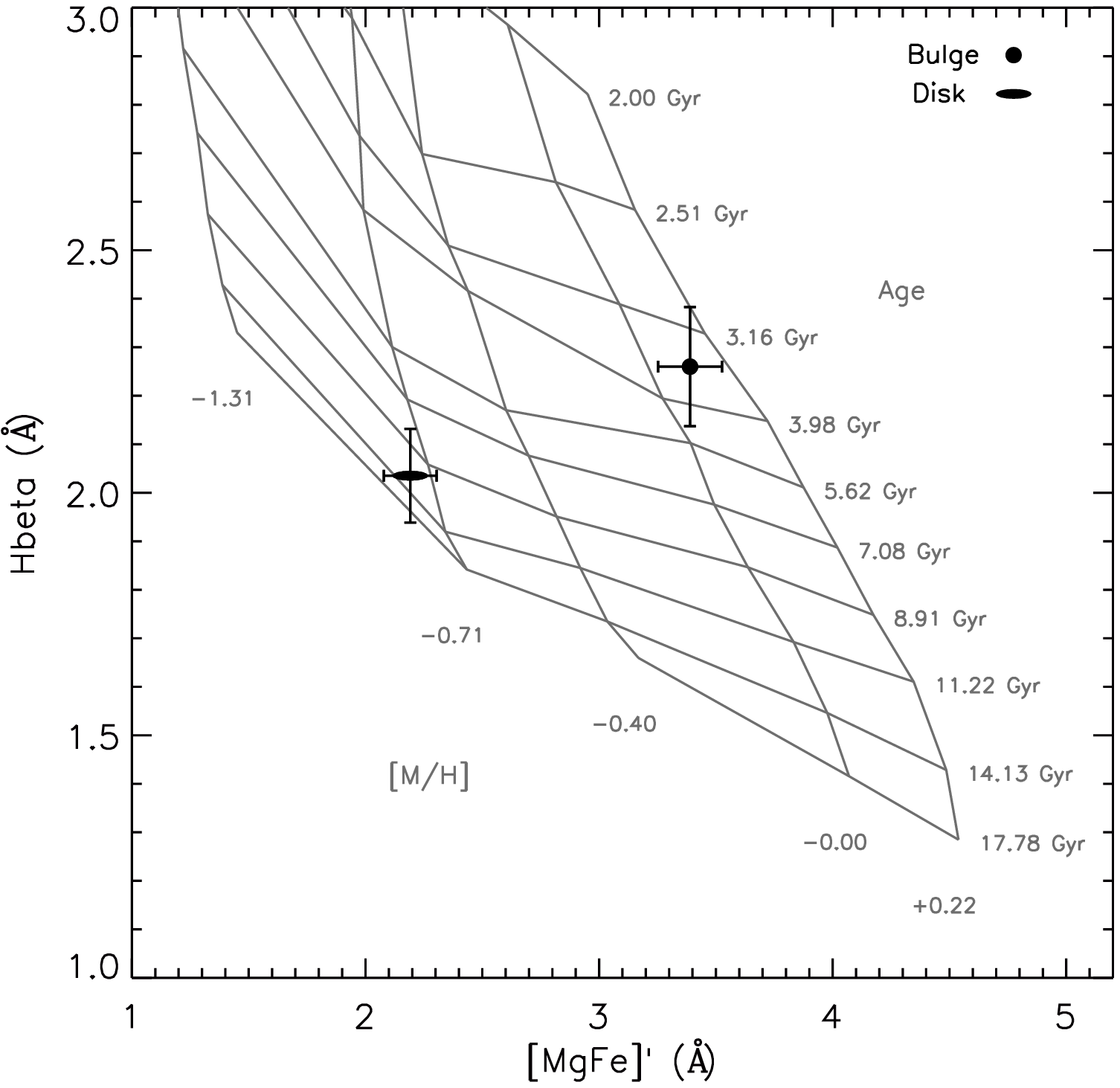}{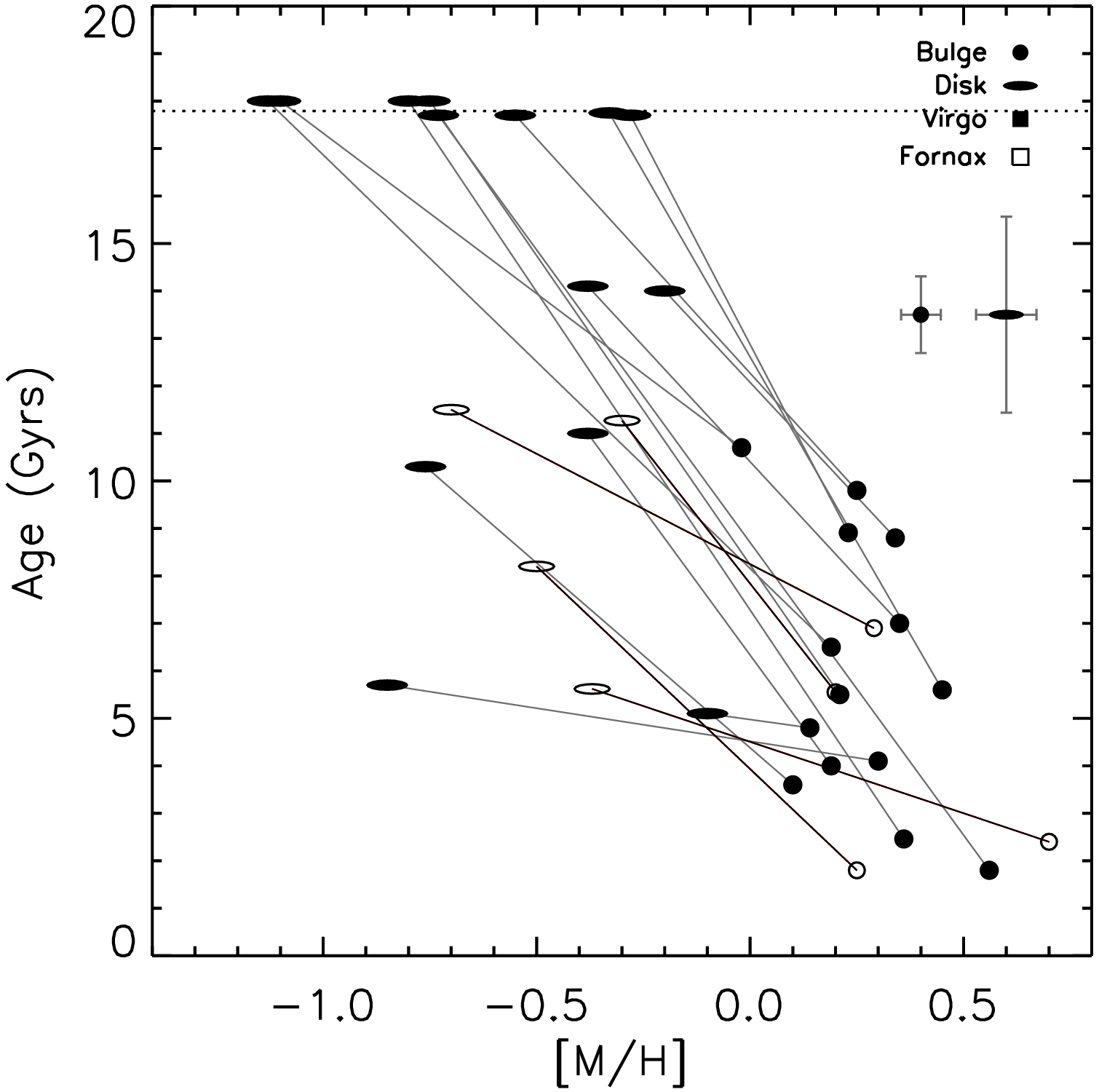}
  \caption{Left: An example SSP model showing relative age and metallicity of 
	    bulge and disk stellar populations in NGC~698.
	  Right: Estimates of the relative ages and metallicities of the bulges 
	  and disks of the Virgo and Fornax Cluster S0s. Note that any disks showing  
	  stellar populations older than the limit of the SSP model (horizontal 
	  dotted line) have been assigned an age of 18~Gyrs.
    \label{Age-Metallicity}}
\end{figure}

The strengths of these absorption features were measured in all the bulge and 
disk spectra, and the combined metallicity index, [MgFe]$'$, calculated. These 
values were then compared to SSP models of \citet{Vazdekis_2010} in order to 
obtain relative light-weighted ages and metallicities for each component. 
An example of an SSP model for the bulge and disk of a galaxy is given in 
Fig.~\ref{Age-Metallicity}, in which it can be seen that the bulge contains
younger and more metal-rich stellar populations than the disk. This trend 
appeared in all the galaxies that could be decomposed, as shown on the right of 
Fig.~\ref{Age-Metallicity}, and suggests that the final star formation event in 
these galaxies occurred within the bulge after the disk had been quenched.

\begin{figure}[t]
\plottwo{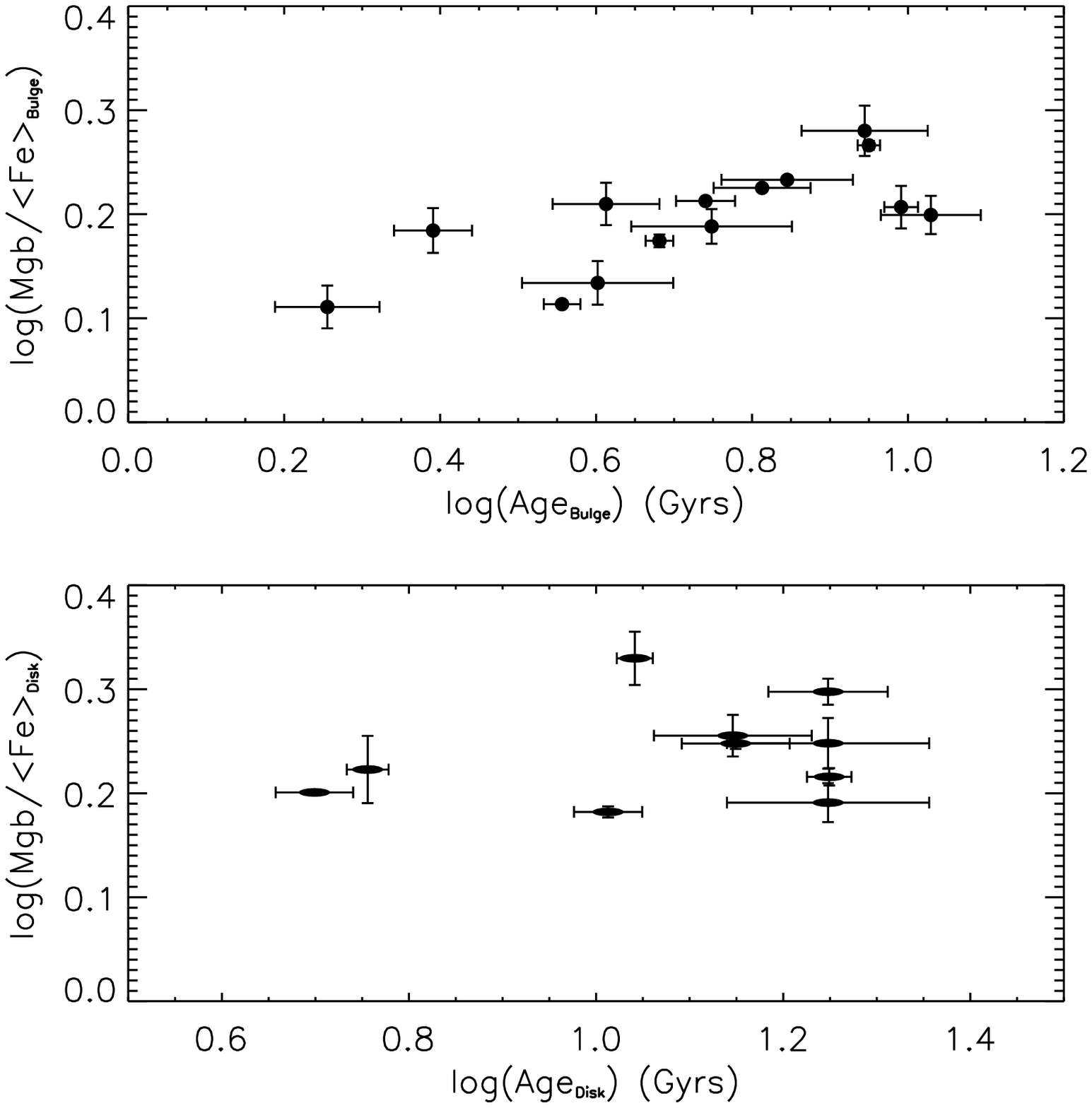}{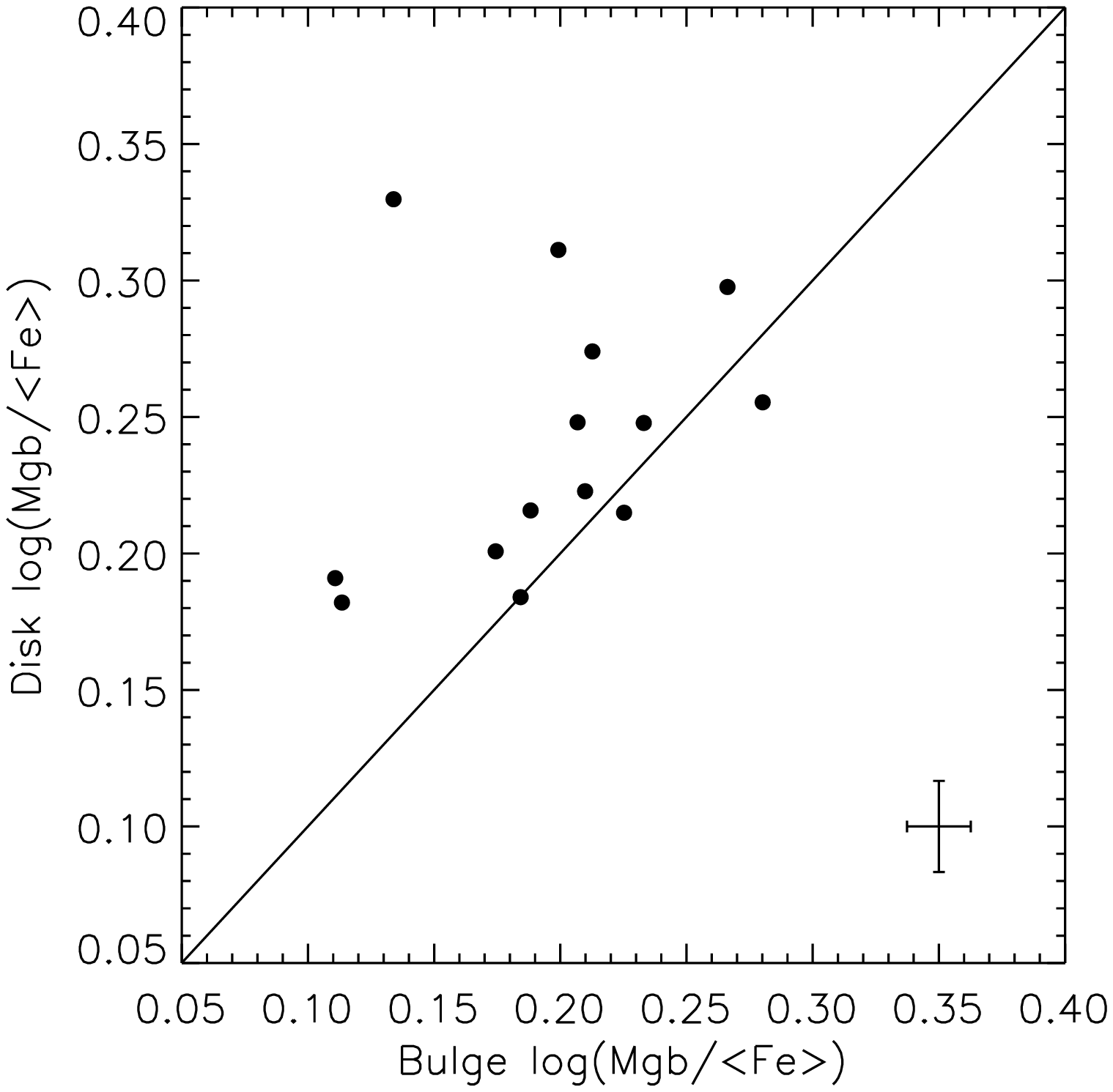}
  \caption{Left: The Mg$/\langle$Fe$\rangle$ ratios of the bulges and disks in the Virgo 
	    Cluster against their ages, showing a clear correlation for the bulges.
	   Right: The Mg$/\langle$Fe$\rangle$ ratios for bulges plotted against those of the 
	    disks, which shows that the bulges are generally more Fe-enriched than 
	    the surrounding disks.
    \label{Figure 2}}
\end{figure}

In order to further interpret this result, the chemical enrichment of the bulges and 
disks was studied. Since SN~II contribute to the Mg abundance 
soon after star formation begins while SN~Ia only enrich the gas with Fe after a 
timescale of around 1~Gyr, the relative abundances of magnesium and iron in 
each component can provide further information on their star formation histories 
\citep{Thomas_2003}.

The Mgb/$\langle$Fe$\rangle$ ratios of the bulges and disks are shown in Fig.~\ref{Figure 2}, 
in which the plot on the top left clearly shows that as the age of the bulge 
increases, it becomes less Fe-enriched. This result suggests that the gas which  
produced the final star formation event in the older 
bulges had undergone a shorter period of star formation than in younger bulges. 
The disks on the other hand (Fig.~\ref{Figure 2}, bottom) show no obvious 
correlation. A possible explanation for this lack of correlation could be that since
the disks are generally quite old, they may have faded significantly since they were 
quenched, and so their measured ages represent their mean global ages as opposed 
to the time since they were quenched. However, the plot on the right of 
Fig.~\ref{Figure 2} shows that the Mgb/$\langle$Fe$\rangle$ ratios from the bulge and disk are 
correlated, with the bulges being generally more Fe-enriched than the disks. This 
result indicates that the final episode of star formation in the bulge is connected 
to the star formation history of the disk.

\section{Conclusions}
We have presented preliminary results of our study of the star formation histories of 
bulges and disks of S0s. We find that the final star formation event within these galaxies 
occurs within the bulges, and that older bulges are less Fe-enriched than their 
younger counterparts. The chemical enrichment of the bulges and disks were also found to 
be related, with the bulges appearing generally more Fe-enriched than the disks. These
results present a scenario for the transformation of spirals to S0s in which the disk 
star formation is quenched due to ram-pressure stripping, and the residual gas is
channelled in towards the centre of the galaxy where it eventually produces 
a final star formation event within the bulge. Such a scenario ties in with the positive 
age and negative metallicity gradients seen in S0s in recent studies by 
\citet{Poggianti_2001}, \citet{Ferrarese_2006}, \citet{SilChenko_2006b} and 
\citet{Kuntschner_2006}, and is starting to reveal a picture consistent with what is seen 
in the distant Universe, but with far more detailed information on the 
transformation histories of individual galaxies.

% Thus, archaeological studies of nearby galaxies are starting to reveal a picture consistent 
% with what is seen in the distant universe, but offers the promise of far more detailed 
% information on the transformation histories of individual galaxies.

\acknowledgements EJ acknowledges support from from the RAS and the STFC.

\bibliography{JohnstonE}

\end{document}